%% file: main.tex
\documentclass[10pt,conference, letterpaper]{IEEEtran}
\IEEEoverridecommandlockouts
\pdfoutput=1
\usepackage{caption}
\usepackage{quantikz}
\usepackage{cite}
\usepackage{amsmath,amssymb,amsfonts}
\usepackage{algorithm}
\usepackage[noend]{algpseudocode}
\usepackage{graphicx}
\usepackage{subcaption}
\usepackage{multirow}
\usepackage{textcomp}
\usepackage{listings}
\usepackage{dcolumn}
\usepackage{bm}
\usepackage{braket}
\usepackage{hyperref}

\usetikzlibrary{decorations.pathmorphing, positioning}
\hypersetup{
    colorlinks = true,
    citecolor = magenta,
    linkcolor = purple
}
\def\BibTeX{{\rm B\kern-.05em{\sc i\kern-.025em b}\kern-.08em
    T\kern-.1667em\lower.7ex\hbox{E}\kern-.125emX}}

\begin{document}
\bstctlcite{IEEEexample:BSTcontrol}

\title{Evaluation of Distimation’s Real-world Performance on a Superconducting Quantum Computer
\thanks{This work is supported by JST Moonshot R$\&$D Grants [JPMJMS226C]. This work acknowledges NICT Quanutm Camp research program.

{\IEEEauthorrefmark{4}These authors contributed equally}}}

\author{
\IEEEauthorblockN{
Hikaru Yokomori\IEEEauthorrefmark{2}\IEEEauthorrefmark{4},
Marii Koyama\IEEEauthorrefmark{1}\IEEEauthorrefmark{4},
Naphan Benchasattabuse\IEEEauthorrefmark{2}\IEEEauthorrefmark{3},
Michal Hajdu\v{s}ek\IEEEauthorrefmark{2}\IEEEauthorrefmark{3},\\
Shota Nagayama\IEEEauthorrefmark{6}\IEEEauthorrefmark{5},
Rodney Van Meter\IEEEauthorrefmark{1}\IEEEauthorrefmark{3}
}

\IEEEauthorblockA{\IEEEauthorrefmark{1}\textit{Faculty of Environment and Information Studies, Keio University Shonan Fujisawa Campus, Kanagawa, Japan}}
\IEEEauthorblockA{\IEEEauthorrefmark{2}\textit{Graduate School of Media and Governance, Keio University Shonan Fujisawa Campus, Kanagawa, Japan}}
\IEEEauthorblockA{\IEEEauthorrefmark{3}\textit{Quantum Computing Center, Keio University, Kanagawa, Japan}}
\IEEEauthorblockA{\IEEEauthorrefmark{6}\textit{Graduate School of Media Design, Keio University, Kanagawa, Japan}}
\IEEEauthorblockA{\IEEEauthorrefmark{5}\textit{mercari R4D, Mercari, Inc., Tokyo, Japan}}\\
\{voy, mia, whit3z, michal, shota, rdv\}@sfc.wide.ad.jp}

\thispagestyle{plain}
\pagestyle{plain}
\maketitle

\begin{abstract}

Quantum state estimation plays a crucial role in ensuring reliable creation of entanglement within quantum networks, yet conventional Quantum State Tomography (QST) methods remain resource-intensive and impractical for scaling.
To address these limitations, we experimentally validate Distimation, a novel distillation-based protocol designed for efficient Bell-diagonal state estimation.
Using IBM Quantum simulators and hardware, we demonstrate that Distimation accurately estimates Bell parameters under simulated and real-world noise conditions, but also demonstrating limitations with operational noise and number of available shots.
Additionally, we simulate an asymmetric-fidelity Bell pair scenario via Measurement-Based Quantum Computation (MBQC) to further validate Distimation under realistic network conditions.
Our results establish Distimation as a viable method for scalable, real-time entanglement monitoring in practical quantum networks.
\end{abstract}


\input{introduction.tex}

\input{background.tex}

\section{Distimation Verification}
\label{sec:distimation-verification}

\input{distimation-protocol.tex}

\section{Real Hardware Distimation Results}
\input{real-distimation.tex}

\section{Asymmetrical Bell Pair Scenario}
\input{asymmetrical-bell-pair-scenario.tex}

\section{Real Device Asymmetrical Scenario}
\input{mbqc-real.tex}

\input{conclusion.tex}

\bibliographystyle{IEEEtran.bst}
\bibliography{IEEEabrv, bibfile}

\end{document}

%% file: introduction.tex
\section{Introduction}
\label{introduction}
Quantum state estimation is essential for verifying and preserving the quality of entanglement in quantum networks~\cite{wehner2018quantum_qn, vanmeter2014quantum_qn,hajdusek2023book}.
In these networks, entangled Bell pairs underpin critical tasks, including quantum teleportation~\cite{bennett1993teleporting}, quantum key distribution~\cite{Bennett1984BB84, ekert1991quantum_qkd, tokyo_qkd, Liao_2017_qkd}, and distributed quantum computing~\cite{BARRAL2025100747, cacciapuoti2020quantum_DQC}.
However, Bell pairs are extremely fragile~\cite{buckley2024bellkat}; their fidelity inevitably deteriorates due to operational noise, transmission, and storage.
The Success of such network applications hinges on the quality of Bell pairs.
Therefore, accurate knowledge of the quantum states that the network distributes is crucial.

Various approaches to state characterization have been developed~\cite{eisert2020quantum}, each with its own trade-off between the resource overhead and the amount of information extracted about the states.
Quantum State Tomography (QST) offers a large information gain, while also being one of the most resource-hungry~\cite{james2001measurement}.
QST fully characterizes the state of an $n$-qubit system at the cost of exponentially many measurement settings, $4^n - 1$ in the worst case.
Furthermore, the state itself is destroyed in the process.
In the context of quantum networks, where generating entangled pairs of qubits is a probabilistic and resource-intensive process, state estimation via QST becomes unsustainable, forming a significant barrier to scaling practical quantum networks.

Addressing these limitations, the recently proposed \textit{Disti-Mator} framework~\cite{maity2023noise,casapao2024distimator} leverages entanglement distillation~\cite{bennett1996purification} as a method for state estimation.
Unlike QST, Disti-Mator infers underlying quantum state parameters directly from the success rates of performed distillation protocols.
Distillation is an integral part of managing errors in quantum repeater networks~\cite{azuma2015allphotonic}, and will therefore be routinely performed as a means to boost entanglement fidelity.
This allows Disti-Mator to estimate the quantum state without the need to pause the usual quantum network operations as is the case for QST.
Furthermore, the Disti-Mator consumes fewer Bell pairs than QST in a fairly broad parameter region.

The Disti-Mator framework currently applies only to Bell-diagonal states, that is states affected by Pauli noise only.
In the case of general noise, it can still be used as a tool to estimate diagonal elements of arbitrary quantum states in the Bell basis but it does not yield any information about the rest of the density matrix.
Furthermore, it was shown in~\cite{casapao2024distimator} that local errors on unitary gates and measurement errors decrease Disti-Mator's performance.

An important open question is the performance of Disti-Mator under real-world conditions, where the above-mentioned issues apply.
However, despite steady experimental progress in entanglement distribution over short~\cite{pompili2021realization,hermans2022qubit,krutyanskiy2023entanglement} and long distances~\cite{zhou2024long,liu2024creation,knaut2024entanglement,krutyanskiy2023telecom}, entanglement distillation has yet to be demonstrated experimentally.
In order to overcome this limitation, we turn to IBM's superconducting quantum computer.
Superconducting qubits have a number of characteristics that make them a suitable platform for evaluating Disti-Mator's real-world performance.
They suffer from amplitude damping channels making the noise model naturally non-Pauli, and
local unitary gates (both single-qubit and CNOT) are innately noisy.
Furthermore, the number of shots per experiment that can be collected is limited, mimicking the low experimental entanglement distribution rates.

We implement the Disti-Mator framework in Qiskit~\cite{qiskit2024} and test its operation using Qiskit's simulator under controlled Pauli noise.
We confirm that our implementation correctly estimates Werner states as well as more general Bell-diagonal states by comparing the output of Disti-Mator with QST. 
Next, we execute Disti-Mator on real hardware and test its performance in the presence of non-Pauli noise and local gate noise.
We find that the output of Disti-Mator differs significantly from QST, with trace distance being 0.26.
We show that this discrepancy is not only due to the amplitude damping channel but also due to limited sample size preventing the Disti-Mator to overcome the effect of local noise~\cite{casapao2024distimator}.
Finally, we investigate the scenario of estimating the quantum states using asymmetric Bell pairs.
This case arises in quantum network operations, where the network distributes the first Bell pair, which then decoheres in quantum memories waiting for successful distribution of the second pair.
We find that this more realistic asymmetric scenario performs slightly better than the symmetric case.

%% file: background.tex
\section{Background}
\label{sec:background}

We now give a brief overview of the Disti-Mator framework~\cite{casapao2024distimator}.
A simple yet useful noise model affecting entangled states is \emph{Pauli noise}, where qubits are assumed to be affected by Pauli $X$, $Y$, and $Z$ errors only.
Applying this noise model to a Bell pair leads to a \emph{Bell-diagonal state}~\cite{horodecki1996separability},
\begin{align}
    \rho_{\text{BD}} & = q_1 \ket{\Phi^+}\bra{\Phi^+} + q_2 \ket{\Phi^-}\bra{\Phi^-} \nonumber\\
    & + q_3 \ket{\Psi^+}\bra{\Psi^+} + q_4 \ket{\Psi^-}\bra{\Psi^-},
    \label{eq:rho_BD}
\end{align}
where $\ket{\Phi^{\pm}}=(\ket{00}\pm\ket{11})/\sqrt{2}$ and $\ket{\Psi^{\pm}}=(\ket{01}\pm\ket{10})/\sqrt{2}$ are the four Bell basis states with $\sum_i q_i=1$.
A special case of Bell-diagonal states is the \textit{Werner state}~\cite{werner1989quantum}, commonly employed to represent isotropic (directionally uniform) depolarization, leading to $q_2=q_3=q_4$.
Werner states can be considered as a mixture of the ideal state, $\ket{\Phi^+}$, and a maximally mixed state,
\begin{equation}
\rho_{\omega} = (1-\omega)\ket{\Phi^+}\bra{\Phi^+} + \frac{\omega}{4}I,\quad \omega \in [0,1],
\end{equation}
where the parameter \(\omega\) quantifies the depolarization level, with \(\omega=0\) indicating a pure maximally entangled state \(\ket{\Phi^+}\), and higher values representing increased isotropic noise.

\begin{figure*}[t]
    \centering
    \begin{tabular}{c c c}
        \begin{quantikz}[column sep=0.14cm, row sep=0.2cm]
        \lstick{$q_0$} & \gate{H} & \ctrl{1} & \gate{I} & \ctrl{2} & \qw       & \qw & \qw \\
        \lstick{$q_1$} & \qw      & \targ{}  & \qw      & \qw      & \ctrl{2}  & \qw & \qw \\
        \lstick{$q_2$} & \gate{H} & \ctrl{1} & \gate{I} & \targ{}  & \qw       & \meter{} & \cw \\
        \lstick{$q_3$} & \qw      & \targ{}  & \qw      & \qw      & \targ{}   & \meter{} & \cw \\
        \end{quantikz}
        &
        \begin{quantikz}[column sep=0.14cm, row sep=0.2cm]
        \lstick{$q_0$} & \gate{H} & \ctrl{1} & \gate{I} & \qw      & \targ{1} & \qw      & \qw & \qw \\
        \lstick{$q_1$} & \qw      & \targ{}  & \qw      & \targ{2}      & \qw  & \qw      & \qw & \qw \\
        \lstick{$q_2$} & \gate{H} & \ctrl{1} & \gate{I} & \qw & \ctrl{-2}      & \gate{H} & \meter{} & \cw \\
        \lstick{$q_3$} & \qw      & \targ{}  & \qw      & \ctrl{-2}  & \qw      & \gate{H} & \meter{} & \cw \\
        \end{quantikz}
        &
        \begin{quantikz}[column sep=0.14cm, row sep=0.1cm]
        \lstick{$q_0$} & \gate{H} & \ctrl{1} & \gate{I} & \gate{S^\dagger} & \qw      & \targ{} & \gate{S} & \qw      & \qw  & \qw\\
        \lstick{$q_1$} & \qw      & \targ{}  & \qw      & \gate{S^\dagger} & \targ{}      & \qw  & \gate{S} & \qw      & \qw & \qw\\
        \lstick{$q_2$} & \gate{H} & \ctrl{1} & \gate{I} & \gate{S^\dagger} & \qw & \ctrl{-2}     & \qw      & \gate{H} & \meter{} & \cw\\
        \lstick{$q_3$} & \qw      & \targ{}  & \qw      & \gate{S^\dagger} & \ctrl{-2}  & \qw      & \qw      & \gate{H} & \meter{} & \cw\\
        \end{quantikz}
    \end{tabular}

    \caption{Quantum circuits for entanglement purification: distillation-(a)(left) detects X and Y errors; distillation-(b)(middle) detects Z and Y errors; and distillation-(c)(right) detects states with X and Z errors. Each circuit begins with two entangled pairs and applies distinct purification logic.}
    \label{fig:pauli-purification}
\end{figure*}
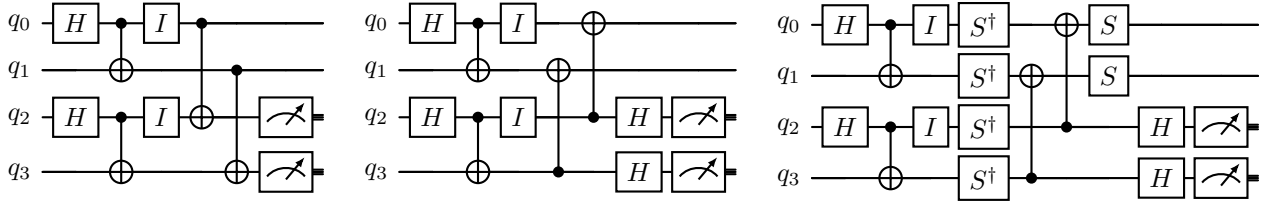

\emph{Entanglement distillation}~\cite{bennett1996purification} is a process that takes two or more Bell pairs of fidelity $F_0$ and probabilistically produces a single Bell pair of fidelity $F_1 \geq F_0$.
The success probability depends on how noisy the initial Bell pairs are.
The main insight behind Disti-Mator is that by estimating the success probability of the distillation process we can estimate the noise parameters.
Figure~\ref{fig:pauli-purification} depicts three distillation protocols that were proposed in~\cite{casapao2024distimator} to estimate Bell-diagonal states in~\eqref{eq:rho_BD}.
The gates labeled $I$ (in Fig.~\ref{fig:pauli-purification}) denote places where Pauli noise is injected into our implementation.
In \textit{Distillation-(a)}, each party performs a CNOT on their two qubits and measures the target (second) qubit in the Z basis.
The procedure is considered successful if both measurements coincide, and the unmeasured qubits are kept.
This protocol detects $X$-type errors ($X$ or $Y$ on a single pair); therefore, a successful distillation increases confidence that no $X$ error has occurred.
\textit{Distillation-(b)} is a variant that reverses the direction of the CNOT gates and uses X basis measurements, allowing detection of $Z$-type error components.
Finally, \textit{Distillation-(c)}\cite{deutsch1996quantum} incorporates local rotations before the CNOTs, enabling the detection of both $X$- and $Z$-type errors simultaneously, though it fails when both errors occur at once (i.e., it cannot detect a $Y$ error).
Bell-diagonal states, as defined in\eqref{eq:rho_BD}, are described by three free parameters and thus require three distinct distillation protocols to fully characterize the Bell-diagonal elements.
In the special case of a Werner state, which is parameterized by a single variable, it suffices to apply any one of the three distillation protocols shown in Fig.~\ref{fig:pauli-purification}.

Previous studies have demonstrated that Distimation provides superior resource efficiency compared to QST for Werner and Bell-diagonal states~\cite{cacciapuoti2020quantum_DQC}. However, it remains unclear whether Distimation maintains its efficacy under more complex and realistic noise conditions typically observed in actual quantum hardware. The original Disti-Mator framework~\cite{casapao2024distimator} emphasizes that its estimation algorithm can tolerate
gate and measurement noise through a bisection-based approach.
However, doing so requires sufficiently large sample sizes to guarantee the desired estimation precision and high success probability, especially under realistic hardware noise.

In this work, we employ a more direct ``inversion method'' for parameter extraction, which uses fewer samples but is more sensitive to gate and measurement errors.
Consequently, our experiments produce larger trace distances compared to the ideal bisection-based approach.
Investigating how to integrate bisection--or other robust post-processing techniques--under practical and realistic sampling constraints remains an important avenue improve accuracy while preserving efficiency, which we leave for future work.

Before proceeding further, we note that for the sake of brevity we will henceforth introduce the verb \emph{to distimate} meaning \emph{to estimate by using distillation}, and the process of estimating via distillation will be referred to as \emph{distimation}.

%% file: distimation-protocol.tex
To verify the Distimation protocol, we translated and implemented it using Qiskit's numerical simulator.
The primary objective is to ensure that Bell-diagonal parameters of noisy Werner and generalized Pauli channels can be accurately estimated through Distimation.
We simulated noisy Bell pairs by configuring specific Pauli error parameters within the quantum circuits.

\subsection{Setup}

\begin{figure}
    \centering
    \includegraphics[width=\linewidth]{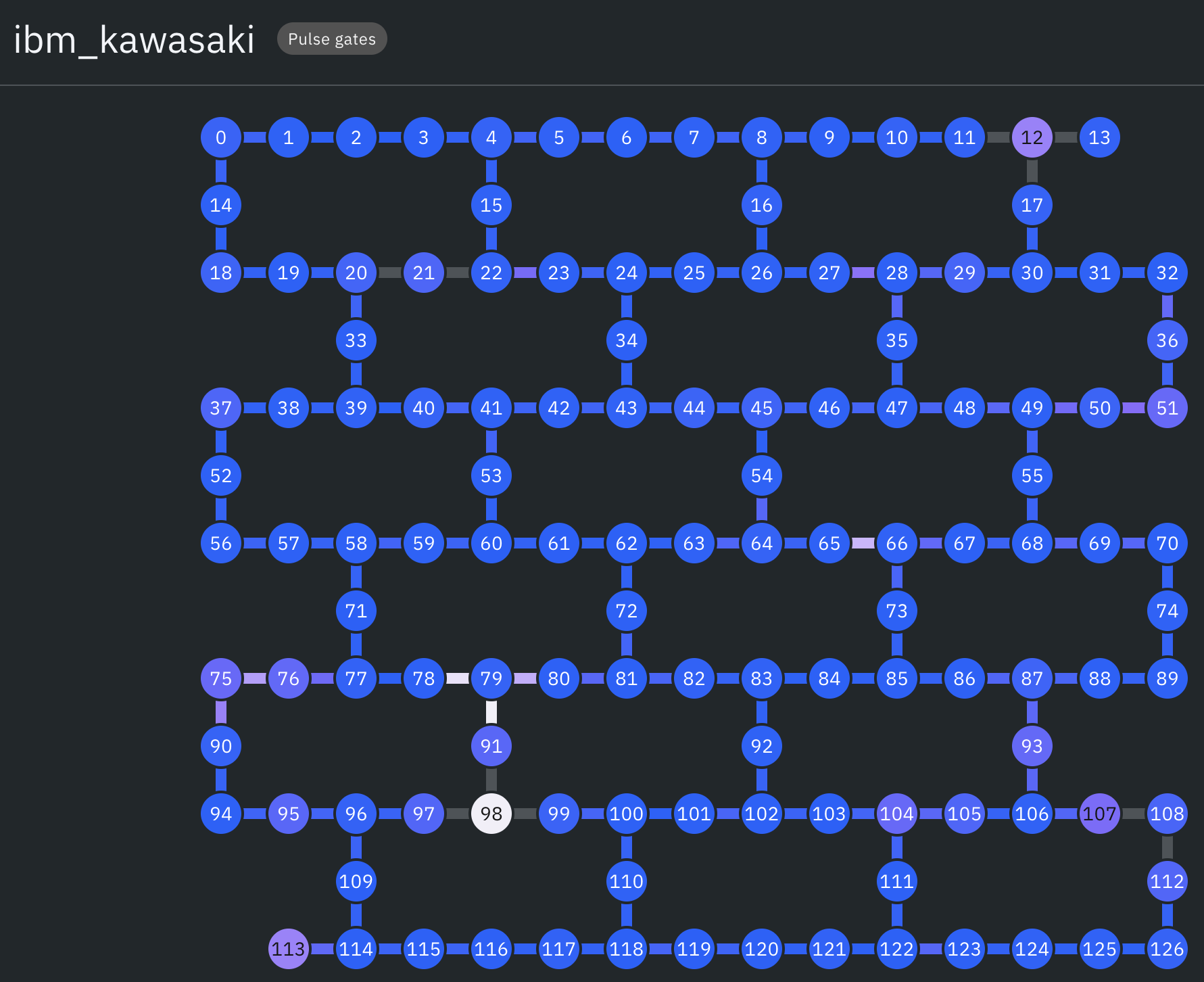}
    \caption{Qubit layout of \texttt{ibm\_kawasaki}, obtained from \texttt{IBM Quantum Platform}. }
    \label{fig:kawasaki-layout}
\end{figure}

We utilized Qiskit's \texttt{AerSimulator} with a \texttt{GenericBackendV2} configuration replicating the qubit layout and coupling map of the \texttt{ibm\_kawasaki} device. We coupled logical qubits: $\{41, 42, 43, 44\}$ to real physical qubits shown in Fig.~\ref{fig:kawasaki-layout}. This allowed us to transpile circuits consistent with actual hardware constraints.

The three two-copy distillation quantum circuits, \textit{Distillation-(a)}, \textit{Distillation-(b)}, and \textit{Distillation-(c)} are shown in Fig.~\ref{fig:pauli-purification}.
Since Bell pairs are symmetric, applying a Pauli noise channel to one qubit of the pair suffices to generate an arbitrary Bell-diagonal mixed state.
In our simulation, we introduced Pauli noise via the identity gates on qubits 0 and 2 using Qiskit's \texttt{NoiseModel} and \texttt{pauli\_error} from \texttt{qiskit\_aer}.

\subsection{Distimation Procedure}

For both Werner and Bell-diagonal state simulations, we first specified the number of Bell pairs and defined relevant error parameters.
The two procedures are described as follows.

\emph{\textbf{Werner Distimation Procedure:}}

\begin{enumerate}
    \item Execute noisy distillation circuits under simulated conditions with noisy Werner state $\rho_W$.
    \item Estimate success probabilities \( \hat{p}^{(i)} \), where \( i \in \{a, b, c\} \), from ``00'' outcomes over total measurements, as defined in the original \textit{Disti-Mator}~\cite{casapao2024distimator}. We note that not including the ``11'' measurement outcomes in \( \hat{p}^{(i)} \), does not affect the Distimation procedure. 
    \item Calculate Werner parameters $\hat{\omega}_i$, where \( i \in \{a, b, c\} \) using 
    \begin{equation}
    \hat{\omega}_{i} = 1 - \sqrt{4\hat{p}^{(i)} - 1}, \quad \text{for } \hat{p}^{(i)} \geq 0.25,
    \label{eq:werner-parameter-calculation}
    \end{equation}
    leading to the estimated density matrix $\hat{\rho}_W$.
    Note that not all distillation protocols are required. Any of the three protocols is capable of successful distimation of a Werner state. We do test all three anyway in order to confirm their correct implementation in Qiskit.
\end{enumerate}

\emph{\textbf{Bell-diagonal Distimation Procedure:}}

\begin{enumerate}
    \item Execute noisy distillation circuits on state $\rho_{\text{BD}}$.
    \item Estimate success probabilities \( \hat{p}^{(i)} \). 
    \item Derive intermediate variables \( \hat{x}_i \) using
    \begin{equation}
    \hat{x}_i = \frac{1}{2}\left(1 + \sqrt{4\hat{p}^{(i)} - 1}\right), \quad \text{for } \hat{p}^{(i)} \geq 0.25.
    \label{eq:bd-intermediate-calculation}
    \end{equation}
    \item Estimate Bell-diagonal parameters via
    \begin{equation}
        \begin{aligned}
        \hat{q}_1 &= \frac{-1 + \hat{x}_1 + \hat{x}_2 + \hat{x}_3}{2}, \\
        \hat{q}_2 &= \frac{1 + \hat{x}_1 - \hat{x}_2 - \hat{x}_3}{2}, \\
        \hat{q}_3 &= \frac{1 - \hat{x}_1 + \hat{x}_2 - \hat{x}_3}{2}, \\
        \hat{q}_4 &= \frac{1 - \hat{x}_1 - \hat{x}_2 + \hat{x}_3}{2},
        \end{aligned}
    \end{equation}
    leading to the estimated Bell-diagonal state $\hat{\rho}_{\text{BD}}$.
\end{enumerate}

\subsection{Evaluation}
\subsubsection{Werner Distimation Evaluation}

We used \(N = 2.7 \times {10^5}\) shots per circuit ( \(5.4 \times 10^5 \) Bell Pairs total) due to Qiskit experimental constraints.
We performed sweeps across Werner states defined by symmetric Bell-diagonal parameters,
\begin{equation}
q = (q_1, q_2 = \frac{1-q_1}{3}, q_3 = q_2, q_4 = q_2).
\end{equation}
with $q_1 \in [0.5, 1.0]$.
We evaluated performance by computing trace distance, \( D(\hat{\rho}_W, \rho_W) = \frac{1}{2}||\hat{\rho}_W - \rho_W||_1, \) between the estimated and the actual Werner states. Fig.~\ref{fig:trace-distance-werner} confirms minimal deviation from zero in trace distance across all Werner parameter, demonstrating effective simulation and verification of the Werner Distimation protocol.
As expected, since the Werner state is parametrized by a single value $\omega$, any of the distillation protocols can be used for effective distimation.

\subsubsection{Bell-diagonal Distimation Evaluation}

\begin{figure}[t]
\centering
\includegraphics[width=0.5\textwidth]{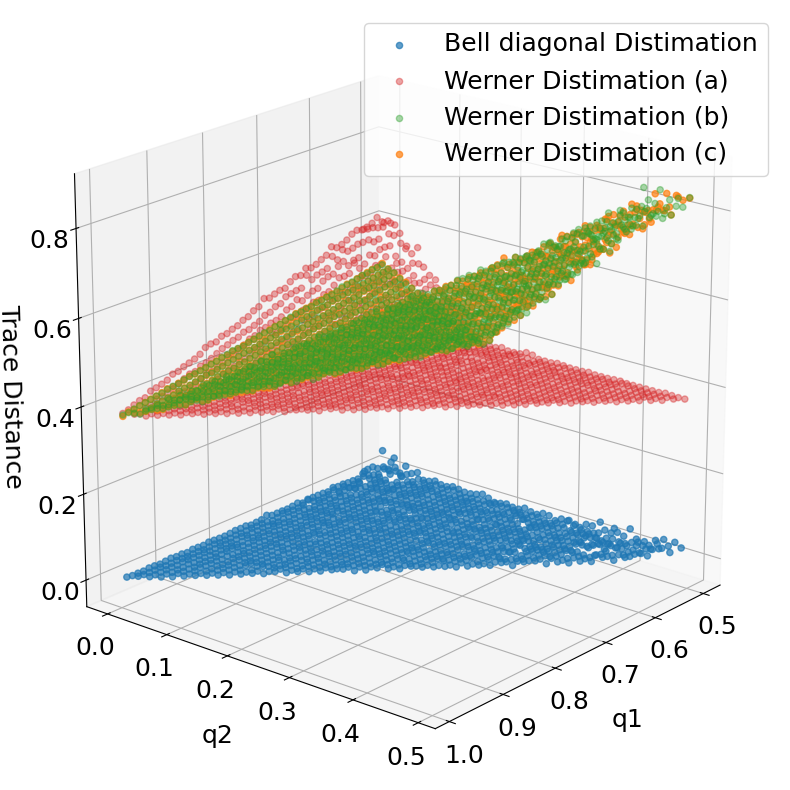}
\caption{Trace distance \( D(\hat{\rho}_{\text{BD}}, \rho_{\text{BD}}) \) across Bell-diagonal parameter space under simulated Pauli noise. Bell-diagonal states cannot be correctly estimated with Werner Distimation procedures alone but require the full Bell-diagonal Distimation.}
\label{fig:trace-distance-bd}
\end{figure}

We used \(N = 9 \times {10^4}\) shots per scenario ( \(5.4 \times 10^5 \) Bell Pairs total). Parameter sweeps were performed for Bell-diagonal states defined by
\begin{equation}
{q} = \left({q_1}, {q_2}, {q_3} = \frac{1 - {q_1} - {q_2}}{2}, {q_4} = \frac{1 - {q_1} - {q_2}}{2}\right)
\label{bell-diagonal-sweep-equation},
\end{equation}
with constraints \( q_1 \in [0.5, 1.0] \) and \( q_2 \in [0, 1 - q_1] \).
The resulting trace distance between the true and estimated Bell parameters across the parameter space under simulated Pauli noise and the Werner state assumption, using the three different distillation protocols, is shown in Fig.~\ref{fig:trace-distance-bd}.
Blue points show that Bell-diagonal distimation procedure is working correctly, as the trace distance is very close to zero across the entire parameter regime.
Trivially, the figure also shows high trace distances for Werner Distimation, confirming its ineffectiveness when applied to non-isotropic Bell-diagonal states.

\begin{figure}[t]
\centering
\includegraphics[width=0.45\textwidth]{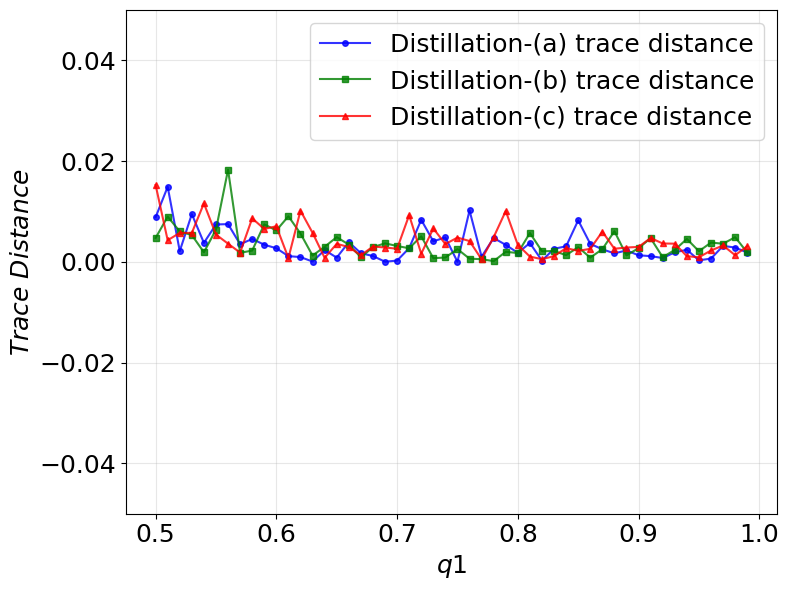}
\caption{Trace distance \( D(\hat{\rho}_W, \rho_W) \) across Werner parameter space under simulated Pauli noise. Minimal variation from zero demonstrates that our distimation of Werner states functions correctly.}
\label{fig:trace-distance-werner}
\end{figure}

Our numerical simulations confirmed that Distimation can estimate Bell-diagonal parameters reliably under controlled
Pauli-noise environments. One key observation from Fig.~\ref{fig:trace-distance-werner} and Fig.~\ref{fig:trace-distance-bd} is the
nearly uniform trace-distance behavior for a wide range of parameter sweeps. This consistency implies that once
Distimation collects enough measurement statistics from the three distillation circuits, its parameter-inference method
renders stable estimates irrespective of whether the underlying noise is isotropic (Werner) or more general (Bell-diagonal).

%% file: real-distimation.tex
\subsection{Werner Distimation Results}

The Werner Distimation method specifically estimates parameters under the Werner state assumption. We evaluated the deviation of the actual quantum state from this assumption and compared Werner parameters estimated from three distinct distillation circuits. 

Recall that the original Disti-Mator protocol~\cite{casapao2024distimator} can, in theory, accommodate depolarizing gate errors if one utilize a bisection-based procedure with sufficiently many samples to ensure accurate parameter bounds.
Unfortunately, for cloud-based runs on the \texttt{ibm\_kawasaki} device, acquiring these large sample counts is impractical.
Hence, we opted for a simpler inversion technique.
While this approach works well for moderate noise, it can yield significant deviations (higher trace distances) once the noise surpasses certain thresholds.
Our data from real hardware runs reflect these trade-offs.
We see that the distillation success probabilities, when combined with limited sampling, do not always suffice to offset gate errors in the final reconstructed state.

We executed the quantum circuits detailed in~Sec.\ref{sec:distimation-verification}---excluding artificially added identity gates---on IBM’s quantum backend \texttt{ibm\_kawasaki}.
Each two-copy distillation circuit was mapped to physical qubits \{41, 42, 43, 44\} to minimize overhead and maintain consistent structure, using \(N = 9 \times 10^4\) shots per circuit (\(5.4 \times 10^5\) Bell pairs total).

Table~\ref{tab:Purification-Result} summarizes the measurement frequencies obtained from these experiments:

\begin{table}[htbp]
    \centering
    \begin{tabular}{| l || c | c | c | c | c |}
         \hline
         & $00$  & $11$   & $10$ & $01$ & Total \\ \hline 
         Distillation-(a) & 23000 & 21984 & 22285 & 22731 & 90000  \\ \hline
         Distillation-(b) & 22782 & 22043 & 22608 & 22567 & 90000 \\ \hline
         Distillation-(c) & 23049 & 22117 & 22472 & 22362 & 90000 \\ \hline
    \end{tabular}
    \caption{Distillation results on \texttt{ibm\_kawasaki}.}
    \label{tab:Purification-Result}
\end{table}

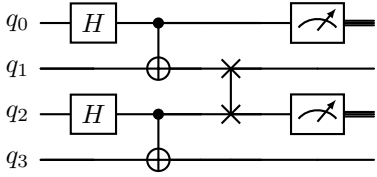
\begin{figure}[t]
    \centering
        \begin{quantikz}[column sep=0.4cm, row sep=0.18cm]
            \lstick{$q_0$} & \gate{H} & \ctrl{1} & \qw & \qw        & \qw   & \meter{}  & \cw \\
            \lstick{$q_1$} & \qw      & \targ{}  & \qw & \swap{1}   & \qw   &   &  \\
            \lstick{$q_2$} & \gate{H} & \ctrl{1} & \qw & \swap{-1} & \qw   & \meter{}  & \cw \\
            \lstick{$q_3$} & \qw      & \targ{}  & \qw & \qw        & \qw   &   &  \\
        \end{quantikz}
    \caption{Quantum circuit for Quantum State Tomography with swap.}
    \label{fig:qst-swap}
\end{figure}

Using these outcomes, we computed success probabilities \(\hat{p}^{(i)}\) and estimated Werner parameters \(\hat{\omega}_i\) via Eq.~\eqref{eq:werner-parameter-calculation} as
\[
\hat{\omega}_a = 0.505, \quad \hat{\omega}_b = 0.181, \quad \hat{\omega}_c = 0.593.
\]
From these Werner parameters, the estimated density matrices \(\hat{\rho}_{\omega_i}\) were constructed using
\[
\hat{q}_1 = 1 - \frac{3\hat{\omega}_i}{4}, \quad \hat{q}_2 = \hat{q}_3 = \hat{q}_4 = \frac{\hat{\omega}_i}{4}.
\]
To assess accuracy, we calculated the trace distance between the Werner Distimated states \(\hat{\rho}_{\omega_{i}}\) and QST-derived actual noisy states \(\rho_{\text{QST}}\).
QST was conducted using \texttt{ParallelExperiment} from \texttt{qiskit\_experiments} on the same qubits.
The logical circuit shown in Fig.~\ref{fig:qst-swap} was mapped to qubits \{41, 42, 43, 44\}.
After Bell state preparation, a subsequent swap operation was applied to align the configuration with that of the distillation circuit at the transpiled level.
Measurements were done on qubits 0 and 2 to estimate Bell pair state corresponding to the logical qubits 0 and 1.
The code prepares 9 circuits, measuring each qubits in different bases\{Z, X, and Y\}.
From the resulting expectation values, it formed a 2-qubit density matrix \(\rho_{\text{QST}}\).

\begin{table}[h!]
\centering
\begin{tabular}{|l|c|}
\hline
\textbf{Distillation Circuit} & \textbf{Trace Distance} \\
\hline
Distillation-(a) & 0.636 \\
Distillation-(b) & 0.417 \\
Distillation-(c) & 0.700 \\
\hline
\end{tabular}
\caption{Trace distance between Werner-distimated and QST states utilizing different distillation method.}
\label{tab:trace_distance_results}
\end{table}

From experiments on the \texttt{ibm\_kawasaki} backend, Table~\ref{tab:trace_distance_results} shows that the success probabilities
for Distillation-(a), Distillation-(b), and Distillation-(c) differ, reflecting the interplay of gate errors, qubit connectivity,
and measurement noise. Interestingly, the estimated Werner parameters $\hat{\omega}_a$, $\hat{\omega}_b$, and
$\hat{\omega}_c$ vary noticeably, suggesting that one particular purification circuit (Distillation-(b) in our case) might
`align' better with the dominant error sources on the hardware. This result again, shows the impracticality of Werner assumption in the real hardware device, confirming it as a toy model. 

\subsection{Bell Diagonal Distimation Results}

Using Table~\ref{tab:Purification-Result}, success probabilities \(\hat{p}^{(i)}\) were computed and used to determine intermediate parameters \(\hat{x}_i\). These were then employed to calculate Bell-diagonal estimates via Eq.~\eqref{eq:bd-intermediate-calculation},
\[
\hat{q} = (0.68,\ 0.07, 0.23,\ 0.02).
\]
From these parameters, the Bell-diagonal state density matrix \(\hat{\rho}_{\text{BD}}\) was formed. Similar to Werner Distimation, the trace distance between \(\hat{\rho}_{\text{BD}}\) and the QST state \(\rho_{\text{QST}}\) was computed as
$D(\hat{\rho}_{\text{BD}}, \rho_{\text{QST}}) = 0.26$.

Moving beyond the symmetric (Werner) noise model, we examined Distimation’s ability to recover asymmetric
Bell-diagonal states on the real hardware. The reconstructed parameters $(\hat{q}_1, \hat{q}_2, \hat{q}_3, \hat{q}_4)$ reveal that the device noise indeed skews the state away from isotropy. The trace distance
between the Bell-diagonal Distimation-derived state $\hat{\rho}_{BD}$ and the QST benchmark is substantially
lower than the average Werner-based distance, indicating that Bell-diagonal Distimation
better captures the real device’s noise profile. This highlights a practical insight: real hardware rarely exhibits strictly
isotropic noise, so adopting the more general Bell-diagonal assumption tends to yield improved modeling fidelity.

\begin{figure}
    \centering
    \includegraphics[width=\linewidth]{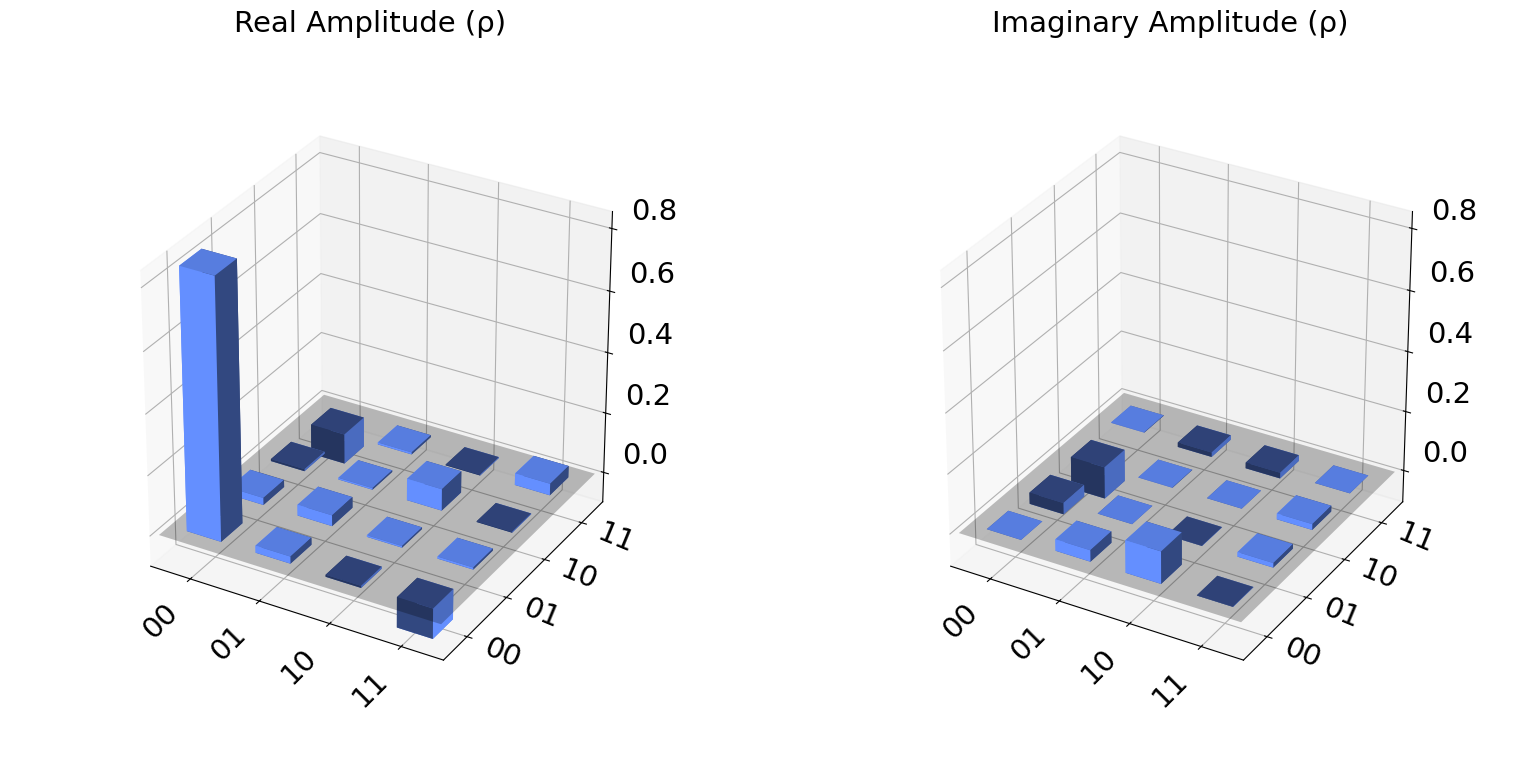}
    \caption{Density Matrix of QST in the Bell Basis, using data from \texttt{ibm\_kawasaki}. Tomography identifies diagonal elements, along with off-diagonal elements. Thus QST directly consumes Bell pairs, the impact on available operational network bandwith is high.}
    \label{bellbasis_qst}
\end{figure}

\begin{figure}
    \centering
    \includegraphics[width=\linewidth]{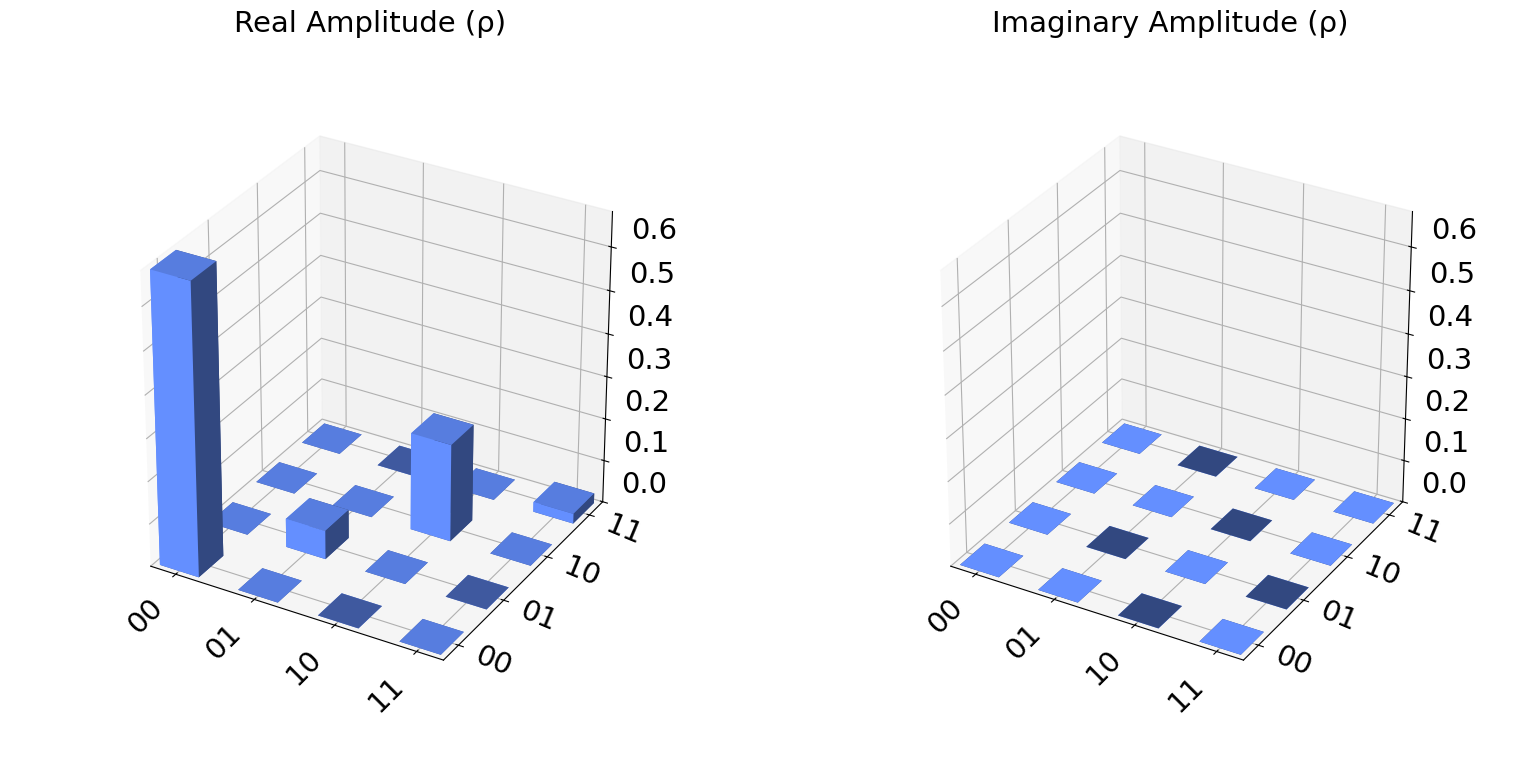}
    \caption{Reconstructed density matrix in Bell basis, using Distimation, of data from \texttt{ibm\_kawasaki}, similar to data in Fig.~\ref{bellbasis_qst}. Note that Distimation only finds the diagonal elements. This reconstruction is valid for many operational tasks and costs are substantially small compared to QST, collecting analyzed data while delivering purified Bell pairs to applications and network uses.}
    \label{bellbasis_estiamte}
\end{figure}

\begin{figure}
    \centering
    \includegraphics[width=\linewidth]{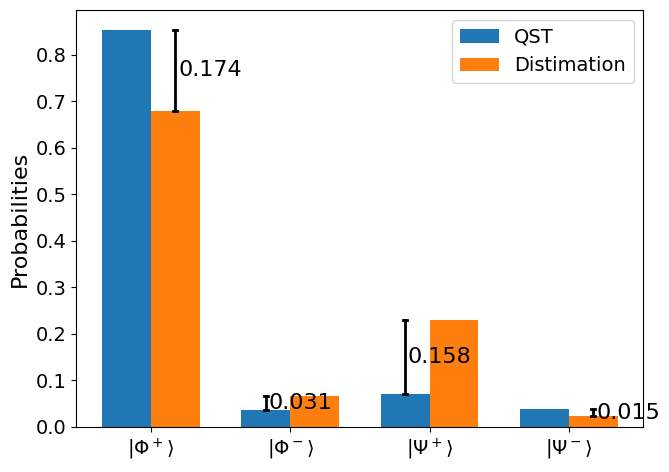}
    \caption{Comparison of QST and Distimation Estimate in Bell Basis}
    \label{fig:bell_basis_real_comparison}
\end{figure}

For a more comprehensive analysis, we compared the true and the estimated state in the Bell basis by examining their Bell-diagonal components.
Specifically, we converted the density matrix $\hat{\rho}_{BD}$ and the state obtained via QST $\rho_{\text{QST}}$ to the Bell basis, as shown in Fig.~\ref{bellbasis_qst} and~\ref{bellbasis_estiamte}.
As one would expect, only the diagonal terms are present in Fig.~\ref{bellbasis_estiamte} since the Bell-diagonal Distimation only reconstructs states diagonal in the Bell basis.
We can see that non-diagonal components are present in Fig.~\ref{bellbasis_qst}.
This is a clear sign of non-Pauli noise affecting the qubits.
Most likely source of these non-diagonal terms is the energy relaxation process. 
To enable a fair comparison aligned with the assumptions of the Bell-diagonal Distimation, we extracted only the diagonal components from both states and labeled them according to their corresponding Bell states, as shown in Fig.~\ref{fig:bell_basis_real_comparison}. Apart from the ideal state $\ket{\Phi^+}$, the most significant discrepancy appears in the population of the $\ket{\Psi^+}$ state, with a difference of 0.158. This deviation indicates that Pauli-X errors are particularly prominent on the quantum device.
Figure~\ref{fig:bell_basis_real_comparison} shows significant discrepancies between the estimated Bell-diagonal parameters $\hat{q}_i$ and the ones recovered via QST.
This is a direct result of the restricted number of samples that we were able to collect from \texttt{ibm\_kawasaki} compounded by the effect of local gate noise.


We reiterate that fully accounting for gate and measurement noise under the Bell-diagonal assumption can require the extensive bisection-based approach, demanding significantly more than our $N=9\times 10^4$ shots. From the complexity analysis in~\cite{casapao2024distimator}, the bisection-based method typically requires $O\!\bigl(\log(1/\delta)/\varepsilon^2\bigr)$ samples for a target error threshold~$\varepsilon$ and failure probability~$\delta$.
In the presence of gate and measurement noise, tighter parameter bounds may push this requirement up by one to two orders of magnitude.
For instance, to achieve target error threshold $\varepsilon \approx 0.01$ (e.g., in trace distance) with high success probability ($\delta \approx 0.01$) can easily demand on the order of $10^6$ total shots \emph{per distillation circuit}, far beyond the $\sim 10^5$ shots we used.
Investigating how to reduce these sampling demands, perhaps through a better post-processing method or via learning theory, remains an important direction of future research.

This trade-off is reflected in our hardware results: under a limited shot budget, the inversion-based estimates are unable to tolerate gate and measurement noise, leading to high trace distances.

Another notable observation is that Distillation-(a) and Distillation-(b) appear to detect contrasting error types, resulting in complementary success probabilities $p^{(a)}$ and $p^{(b)}$.
In a real system, $X$, $Y$, or $Z$ errors do not always occur at identical rates.
Distillation-(a), focusing on $X$-type error detection, might report lower success rates on hardware dominated by $Z$-type errors, while Distillation-(b) may exhibit the opposite behavior.
Distillation-(c)’s rotations are intended to detect either $X$ or $Z$ errors while letting $Y$ errors go undetected, which is useful when the noise is biased towards those two Pauli noise, but in practice, but in real systems as seen above, the noise is not biased towards $X$ or $Z$ alone.
Thus, the estimated Bell-diagonal distribution $(q_1,q_2,q_3,q_4)$ can serve as a valuable signature for diagnosing which error channels dominate on a particular device.
Over time, repeated Distimation cycles can thus form a real-time profile of the device’s error environment, complementing or reducing the need for standard tomography.

%% file: asymmetrical-bell-pair-scenario.tex
In practical quantum network implementations, entangled Bell pairs frequently experience asymmetrical fidelity due to sequential generation and storage requirements. Particularly, in entanglement distillation protocols, multiple Bell pairs must be independently generated and stored in quantum memory until all required pairs become available for distillation. This waiting period introduces asymmetrical decoherence, disproportionately affecting earlier-generated pairs. To realistically evaluate the Distimation protocol under these practical network conditions, we implemented a Measurement-Based Quantum Computation (MBQC)~\cite{raussendorf2001oneway} scheme utilizing a one-dimensional (1D) cluster state on Qiskit's \texttt{AerSimulator}.

\subsection{Decoherence from Sequential Bell Pair Generation}

Quantum network operations often rely on probabilistic, sequential Bell pair generation. Earlier-generated Bell pairs must remain idle in quantum memory while subsequent pairs are generated~\cite{khatri2021memory}. During this waiting period, decoherence mechanisms disproportionately degrade earlier-generated pairs compared to recently created ones, resulting in asymmetric fidelity distributions~\cite{buckley2024bellkat}. Consequently, realistic quantum networks inherently exhibit asymmetric fidelities across stored Bell pairs, challenging reliable entanglement verification.

\subsection{MBQC-Based Bell Pair Generation via 1D Cluster States}

\begin{figure}[t]
\centering
\begin{tikzpicture}
\def\myvdots{\ \vdots\ }
  \node (qc1) at (0,4) {
    \begin{quantikz}[column sep=0.2cm, row sep=0.2cm]
    \lstick{$q_0$} & \gate{H} & \ctrl{1} & \qw \\
    \lstick{$q_1$} & \qw      & \targ{}  & \qw
    \end{quantikz}
  };

\node (qc2) at (0,0) {
\begin{quantikz}[column sep=0.1cm, row sep=0.1cm]
    \lstick{$q_0$}      & \gate{H}      & \ctrl{1}  & \qw            & \qw               & \qw           & \gate{Z}          & \push{\cdots}\qw & \gate{Z}       & \gate{H}     &\qw\\
    \lstick{$q_1$}      & \gate{H}      & \control \qw & \ctrl{1}    & \meter{}          & \cwbend{6}\setwiretype{}    & \setwiretype{}    & \setwiretype{}   & \setwiretype{} & \setwiretype{}&\\  
    \lstick{$q_2$}      & \gate{H}      & \ctrl{1}  & \control \qw   & \meter{}          & \cw\setwiretype{}           & \cwbend{-2}\setwiretype{}       & \setwiretype{}   & \setwiretype{} & \setwiretype{}&\\ 
    \lstick{\myvdots}   & \setwiretype{} & \vdots{} & \setwiretype{} & \setwiretype{}   & \setwiretype{} & \setwiretype{}   & \setwiretype{}    & \setwiretype{} & \setwiretype{}&\\
    \lstick{}           & \setwiretype{} & \vqw{1}  & \setwiretype{} & \setwiretype{}   & \setwiretype{} & \setwiretype{}   & \setwiretype{}    & \setwiretype{} & \setwiretype{}&\\ 
    \lstick{$q_{n-2}$}  & \gate{H}      & \ctrl{-1} & \ctrl{1}       & \meter{}          & \cw\setwiretype{}           &\cw\setwiretype{}               & \cwbend{2}\setwiretype{}        & \setwiretype{} & \setwiretype{}&\\ 
    \lstick{$q_{n-1}$}  & \gate{H}      & \ctrl{1}  & \control \qw   & \meter{}          & \cw \setwiretype{}          & \cw\setwiretype{}              & \cw\setwiretype{}               & \cwbend{-6}\setwiretype{}    & \setwiretype{}&\\ 
    \lstick{$q_n$}      & \gate{H}      & \control \qw & \qw         & \qw               & \gate{Z}      & \push{\cdots}\qw  & \gate{Z}          & \qw           & \qw           &\qw\\
\end{quantikz}
  };

  \draw[->, thick] (qc1.south) -- (qc2.north) node[midway, above]{}; 

\end{tikzpicture}
\caption{Transformation from a 2-qubit 
 entangled circuit to a generalized $n$-qubit entangled circuit, utilizing MBQC scheme.}
\label{fig:mbqc-bellpair-circuit}
\end{figure}
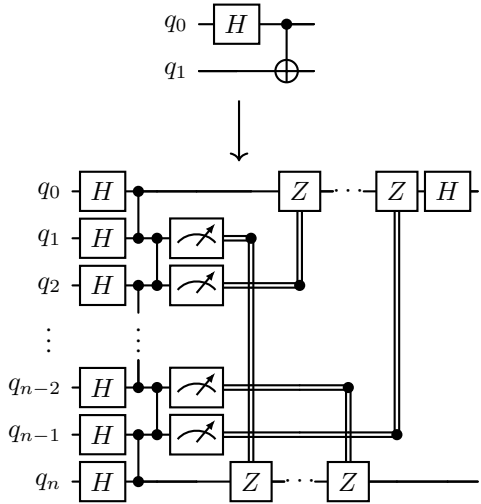

To accurately simulate this practical scenario, we employed MBQC with a 1D cluster state configuration, as illustrated in Fig.~\ref{fig:mbqc-bellpair-circuit}.
Initially, Hadamard gates are applied to all qubits to prepare them in the $\ket{+}$ state, forming the foundation for the cluster state.
Subsequently, controlled-$Z$ (CZ) gates are applied between adjacent qubits, entangling them into a linear cluster.
Single-qubit X basis measurements are performed on intermediate qubits (qubits $q_1$ to $q_{n-1}$) propagate the entanglement, effectively projecting the large resource state down to a single Bell pair located at the cluster endpoints ($q_0$ and $q_n$).
Depending on measurement outcomes, feed-forward operations (Z gates) are applied conditionally to endpoint qubits correcting the projected state into a known target Bell pair.
Constrained by the IBM quantum backend qubit topology, our MBQC-based Bell pair utilized a total of 10 qubits.
This approach of creating entanglement by measuring a cluster state is analogous to proposals for all-photonic quantum repeaters~\cite{azuma2015allphotonic}, which use such states instead of quantum memories. 

To verify the correctness of our MBQC-generated Bell pairs, we conducted QST on qubits $q_0$ and $q_n$ after completing the MBQC protocol on a Qiskit simulator.
The resulting density matrix $\rho_{\text{MBQC}}$ was reconstructed, and the fidelity with the ideal Bell state $\ket{\Phi^+}$ was computed.
Without additional noise, the fidelity was verified to be exactly 1, confirming correct implementation.

\subsection{MBQC-based Distimation Verification}
To verify the Distimation protocol using the MBQC-generated Bell pairs, we performed simulations using the purification circuits illustrated earlier in Fig.~\ref{fig:pauli-purification}, except inplementing the first Bell pair, qubits 0 and 1 using the MBQC scheme from Fig~\ref{fig:mbqc-bellpair-circuit}. 

Similar to the evaluation in Sec.~\ref{sec:distimation-verification}, Pauli error parameter was swept as~\eqref{bell-diagonal-sweep-equation}, with constraints \( q_1 \in [0.5, 1.0] \) and \( q_2 \in [0, 1 - q_1] \) (increments of 0.01).

The resulting trace distance between true and estimated Bell parameters across the parameter space under simulated noise, showed minimal variation in trace distance across all Bell diagonal parameter sweeps for Bell diagonal Distimation, similary to Fig~\ref{fig:trace-distance-bd}. This demonstrates viability of Distimation under MBQC scheme implementation.

%% file: mbqc-real.tex
Our experiment involved generating and verifying MBQC-based Bell pairs on specific qubit layouts and subsequently performing Distimation to estimate the underlying Bell-diagonal parameters. We explicitly tested scenario, characterized by distinct physical qubit endpoints, to verify consistency across different hardware qubit locations and connectivity paths.

\subsection{Experimental Setup and Procedure}

\begin{figure}[ht]
  \centering
  \begin{tikzpicture}[scale=0.6, every node/.style={scale=0.9}]
    \tikzstyle{qubit}=[circle, draw, minimum size=14pt, inner sep=0pt]

    \node[qubit, fill=yellow!30] (q41) at (0,2) {};
    \node[above=4pt of q41] {\textbf{41}};
    
    \node[qubit, fill=yellow!30] (q42) at (2,2) {};
    \node[above=4pt of q42] {\textbf{42}};
    
    \node[qubit, fill=yellow!30] (q43) at (4,2) {};
    \node[above=4pt of q43] {\textbf{43}};
    
    \node[qubit, fill=yellow!30] (q44) at (6,2) {};
    \node[above=4pt of q44] {\textbf{44}};
    
    \node[qubit, fill=yellow!30] (q45) at (8,2) {};
    \node[above=4pt of q45] {\textbf{45}};
    
    \node[qubit, fill=yellow!30] (q54) at (8,0) {};
    \node[above=4pt of q54] {\textbf{54}};
    
    \node[qubit, fill=yellow!30] (q64) at (8,-2) {};
    \node[above=4pt of q64] {\textbf{64}};
    
    \node[qubit, fill=yellow!30] (q63) at (6,-2) {};
    \node[above=4pt of q63] {\textbf{63}};
    
    \node[qubit, fill=yellow!30] (q62) at (4,-2) {};
    \node[above=4pt of q62] {\textbf{62}};
    
    \node[qubit, fill=yellow!30] (q61) at (2,-2) {};
    \node[above=4pt of q61] {\textbf{61}};
    
    \node[qubit, fill=blue!30] (q60) at (0,-2) {};
    \node[above=4pt of q60] {\textbf{60}};
    
    \node[qubit, fill=blue!30] (q53) at (0,0) {};
    \node[above=4pt of q53] {\textbf{53}};

    \draw[thick] (q41)--(q42)--(q43)--(q44)--(q45);
    \draw[thick] (q60)--(q61)--(q62)--(q63)--(q64);

    \draw[thick] (q45)--(q54)--(q64);
    \draw[thick, blue] (q41)--(q53)--(q60);
    \draw[thick] (q43)--++(0,0.8);
    \draw[thick] (q62)--++(0,-0.8);

    \draw[thick] (q41)--++(-1,0);
    \draw[thick] (q45)--++(1,0);
    \draw[thick] (q60)--++(-1,0);
    \draw[thick] (q64)--++(1,0);

    \colorlet{cnotcolor}{black!70!blue}

    \draw[cnotcolor, line width=1.8pt, opacity=0.9]
      (q41) .. controls (-1.5,2.5) and (-1.5,-0.5) .. (q53);
    \fill[cnotcolor, opacity=0.9] (q41) circle (3.2pt); 
    \node[draw, circle, minimum size=10pt, inner sep=0pt,
          line width=1.2pt, cnotcolor, fill=white, fill opacity=0.6,
          text=cnotcolor, text opacity=0.9]
          at (q53) {\textbf{\large X}};

    \draw[cnotcolor, line width=1.8pt, opacity=0.9]
      (q61) .. controls (1.5,-3) and (-1.5,-3) .. (q60);
    \fill[cnotcolor, opacity=0.9] (q61) circle (3.2pt); 
    \node[draw, circle, minimum size=10pt, inner sep=0pt,
          line width=1.2pt, cnotcolor, fill=white, fill opacity=0.6,
          text=cnotcolor, text opacity=0.9]
          at (q60) {\textbf{\large X}};

  \end{tikzpicture}
  \caption{
    Qubits involved in the MBQC-based Distimation experiment are shown on a cropped version of the device’s topology for clarity.
    The layout highlights the relevant qubits with enlarged nodes, annotated connections, and stylized CNOT gates (dark blue lines) to indicate the CNOT used in Distimation circuit.
    }
  \label{fig:kawasaki}
\end{figure}
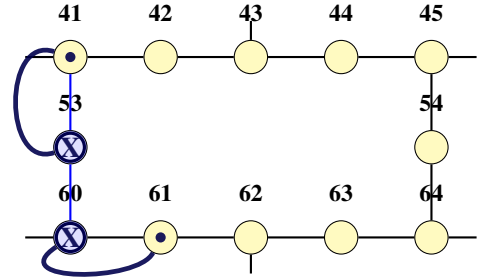

Experiment was executed on the IBM Quantum backend \texttt{ibm\_kawasaki}(Fig~\ref{fig:kawasaki-layout}). 
Due to device coupling constraints , we carefully mapped the MBQC-based Bell pair generation circuits onto a fixed set of physical qubits, $\{41, 42, 43, 44, 45, 54, 64, 63, 62, 61, 60, 53\}$, as shown in Fig~\ref{fig:kawasaki}. 

The initial undistilled Bell pair (qubits $q_0$, $q_9$) generated by the MBQC method was directly characterized via the Distimation protocol. The MBQC scheme connected qubits 41 to 61, simulating the decoherence-induced asymmetry commonly observed in quantum repeaters. This was done by measuring the short undistilled Bell Pair qubits 53 and 60. Diagram of experiment is shown in Fig~\ref{fig:kawasaki}. The Bell-diagonal parameters estimated through Distimation circuits (\textit{Distillation-(a)}, \textit{(b)}, and \textit{(c)}) were then compared to the QST-derived state to assess the accuracy of the Distimation protocol.

The experiment proceeded as follows:

\begin{enumerate}
    \item A linear 10-qubit cluster state was initialized, followed by single-qubit measurements with adaptive corrections ($Z$ gates) to project the state deterministically into a Bell pair at the respective endpoints.
    \item QST was then performed on the endpoint qubits to obtain the experimental density matrix $\rho_{\text{QST}}$.
    While for MBQC-Distimation, we perform measurements on the Bell pairs outside of the chain (qubits 53 and 60).
    Note that no SWAP operation was required, unlike the non-MBQC case shown in Fig.~\ref{fig:qst-swap}, with our choice of MQBC chain location.
    \item Subsequently, the Distimation protocol was implemented via the three distillation circuits (\textit{Distillation-(a)}, \textit{(b)}, and \textit{(c)}) to estimate the Bell-diagonal parameters. Each circuit was executed with $N = 9\times 10^4$ shots per scenario to maintain statistical reliability.
\end{enumerate}

\subsection{Experimental Results and Evaluation}

Measurement outcome frequencies obtained from the quantum hardware experiment is presented in Table~\ref{tab:real-mbqc-results}. Using these outcomes, success probabilities ($\hat{p}^{(i)}$) for each distillation procedure were computed, allowing estimation of Bell-diagonal parameters ($\hat{q}_i$). These parameter estimates were then used to reconstruct their respective density matrices ($\rho_{\text{QST}}$, $\hat{\rho}_{\text{BD}}$).

To evaluate the accuracy and practical utility of Distimation under realistic device conditions, we calculated trace distances between the experimentally reconstructed Distimation states and the QST-measured reference states ($\rho_{\text{QST}}$). The computed trace distances was 0.163.

\begin{table}[htbp]
\centering
\begin{tabular}{|l|c|c|c|c|c|}
\hline
\textbf{Circuit } & `00' & `11' & `10' & `01' & Total \\ \hline
Distillation-(a) & 28699 & 25767 & 15873 & 19661 & 90000 \\ \hline
Distillation-(b) & 28402 & 26201 & 16359 & 19038 & 90000 \\ \hline
Distillation-(c) & 28604 & 26244 & 16231 & 18921 & 90000 \\ \hline
\end{tabular}
\caption{Real-device measurement frequencies for MBQC-generated Bell pairs.}
\label{tab:real-mbqc-results}
\end{table}

 Our results illustrate how this asymmetry can significantly influence the distillation success
probabilities $p^{(i)}$. From Table~\ref{tab:real-mbqc-results} and the ensuing trace distance analysis, we observe that even though the
earliest (and thus noisiest) Bell pairs degrade more quickly, Distimation can still recover meaningful estimates of the
resulting Bell-diagonal parameters. Specifically, the presence of local $CZ$ gates and adaptive measurements in MBQC
introduces an additional layer of gate noise, but the distillation success rates remain sufficient to produce stable parameter
inferences. If the pairs are heavily imbalanced (one of them was stored
far longer than the other), the net fidelity gains from distillation may be smaller, impacting the Distimation reconstruction. The MBQC-based experiments demonstrate Distimation’s robustness under more realistic quantum networking
conditions, where sequential generation and storage of Bell pairs produce asymmetric fidelities.

Similar to our earlier discussion, the MBQC-based Distimation inherits the same sampling constraints. Although
Disti-Mator can, in theory, handle gate noise with bisection-based estimation, that level of rigor is beyond our current
shot budget. This limitation explains why the trace distance of $\hat{\rho}_{BD}$ is 0.163, which remains larger than one might expect if
fully implementing the noise-tolerant version. A future effort is to refine our post-processing to reduce the sample
overhead.

Nonetheless, these findings reinforce Distimation’s key advantage: it exploits the exact same operations that
would be done for entanglement purification, so no extra quantum resources are required. Even in an MBQC-based
entanglement-distribution protocol, Distimation seamlessly piggybacks on the existing CNOT-like and measurement
operations to yield real-time estimates. This shows promise for large-scale quantum networks, where advanced
photonic-based or cluster-state-based repeater strategies will almost certainly generate entangled pairs at slightly
different times. Having a passive, distillation-derived characterization tool in such scenarios could be essential for
monitoring network reliability and optimizing resource usage in dynamic, multi-hop quantum communication setups.

%% file: conclusion.tex
\section{Conclusion}
\label{sec:conclusion}

We have validated Distimation for estimating Bell-diagonal states using distillation protocols on IBM Quantum hardware. Both simulations and real-device tests confirm that Distimation can reconstruct Bell-diagonal parameters for Werner and general Bell-diagonal states, even in scenarios where noise is asymmetric across multiple Bell pairs.
Furthermore, the MBQC-based experiments confirm that Distimation is well suited to realistic quantum network settings, in which entangled pairs have varying fidelities.

Finally, we emphasize that although our implementation successfully demonstrated Distimation on real hardware, it did
not employ the bisection-based sample-bounding procedure advocated by the original Disti-Mator framework~\cite{casapao2024distimator}.
Achieving true noise tolerance may require larger sample sizes than were
feasible here. Exploring ways to reduce these sampling demands—whether via partial bisection,
or advanced classical post-processing represents an exciting direction.
Improvements along this line can potentially
shrink the trace distances we observed and bring real-time entanglement monitoring closer to theoretical ideals.

Further reducing overhead remains critical for the next phase of Distimation development. Beyond this, incorporating \emph{time-varying} Bayesian inference techniques to dynamically adapt parameter estimates under drifting hardware noise will be a key direction. Moreover, investigating Distimation with advanced purification strategies beyond two-copy protocols~\cite{fujii09}, and extending it to multi-qubit or multi-party entanglement scenarios, presents an exciting avenue for scaling the approach. As quantum technologies evolve within the Noisy Intermediate-Scale Quantum (NISQ) era~\cite{Preskill_2018}, transitioning toward Fault-Tolerant Quantum Computing (FTQC), error correction schemes~\cite{Muralidharan2016} should also be integrated into the protocol design.

We anticipate that Distimation will become a useful tool for quantum network operators, enabling accurate, near real-time entanglement diagnostics by leveraging the same resources used for distillation, thereby streamlining state estimation in practical quantum systems.